\newcommand{\be}{\begin{equation}}
\newcommand{\ee}{\end{equation}}
\newcommand{\bea}{\begin{eqnarray}}
\newcommand{\eea}{\end{eqnarray}}
\newcommand{\bc}{\begin{center}}
\newcommand{\ec}{\end{center}}
\documentstyle[preprint,aps]{revtex}
\begin{document}
\draft
\preprint{Alberta-Thy-10-95}
\title{Coarse-Graining and Renormalization Group  in the Einstein
Universe}
\author{ALFIO BONANNO}
\address{Canadian Institute for Advanced Research Cosmology
Program,\\
Theoretical Physics Institute, University of  Alberta,\\
Edmonton, Alberta, Canada T6G 2J1\\
and\\
Institute of Astronomy, University of Catania\\
Viale Andrea Doria 6, 95125 Catania, Italy}
\maketitle
\begin{abstract}
The Kadanoff-Wilson renormalization group approach  for a scalar
self-interacting field theory generally coupled with gravity is
presented.  An
average potential that monitors the fluctuations of  the blocked
field in
different scaling regimes
 is constructed in a nonflat background and
explicitly computed  within the loop-expansion approximation
for an
Einstein
universe.  The curvature turns out to be dominant in setting the
crossover
scale from a double-peak and a symmetric distribution of the block
variables.
The evolution of all  the coupling constants generated by the
blocking
procedure is examined:
the renormalized trajectories agree with the standard perturbative
results for
the relevant vertices near the ultraviolet fixed point, but new
effective
interactions between gravity and matter are present.  The flow of the
conformal
coupling constant is therefore analyzed in the improved scheme and
the infrared
fixed point  is reached for arbitrary values of the renormalized
parameters. \\
PACS number(s): 04.62+v, 11.10.Hi, 05.40+j

\end{abstract}
\baselineskip 14pt
\section{Introduction}

An outstanding issue that has still to be addressed within the
theories of a
quantum field in curved spacetime, concerns the understanding of the
infrared
domain of the models. While powerful techniques (see \cite{bd} for a
review)
clarified the renormalizability properties and the structure of the
effective Lagrangians describing the physics of the
ultraviolet fixed point, the local character of the methods
applied is a serious limitation to studying the infrared scaling
behavior
of a quantum field in a non-trivial topology.

When the system strongly fluctuates over all the scales, the
global topology and, especially in cosmology,  the
dynamic nature of the background spacetime, play a dominant role in
determining long-range features of the system. In this case the
physics is characterized by the scaling around the infrared fixed
point, and
the usual set of renormalizable operators may not be enough to
build the low-energy effective Lagrangian.

To describe the change in the physics as the cutoff is lowered, we
need to follow  the evolution of  all the coupling constants,  both
relevant and nonrelevant, as we approach
the infrared region. The evolution of the latter slows down when we
leave the
UV fixed point towards the infrared. However they
need to be followed
as well, since they may have a different scaling
law in the deep low-energy
domain.  In particular
the classification of the operators close to the UV or the IR fixed
point can
be radically different.
For instance, in  flat space,
the ultraviolet and the infrared are separated by a crossover region
at the
mass gap in a massive theory. Below the mass scale fluctuations are
strongly
damped and the scaling around the infrared  fixed point is monitored
only by
the mass term. But if the theory is massless, the infrared sector is
plagued by
divergences indicating that  a new set of IR relevant operators is
needed. The
actual form of this scaling operators can be very complicated in
terms of
local field variables because, just at the IR fixed point, the physics
is essentially nonlocal
and therefore the global properties of the
spacetime affect dramatically the long range behavior
of these models.

In a curved spacetime the theory presents a new characteristic
length, the local radius of
curvature that usually couples with the field as a new,
nonhomogeneous,
mass gap
$\frac{1}{2}\xi R(x)\phi^2(x)$ where $R(x)$ is the scalar of
curvature. For
this
reason fluctuations of the field on regions that are comparable with
the local
radius of the spacetime are damped and eventually, at energies well
below the
``gravitational" mass contribution, an infrared fixed point can form
even for
models that are massless at the bare level. In this case the
crossover
region is determined by the interplay between matter and gravity. In
particular
one expects that in the strong curvature regime the gravitational
contribution
is the leading one in monitoring the scaling properties of the
theory and in
determining the location of the infrared fixed point.

The Kadanoff-Wilson \cite{kw} realization of the renormalization
group
approach used in statistical mechanics allows one to follow the
evolution of an infinite number of coupling constants as the
cutoff changes.
The natural framework for describing fluctuations at different
relevant scales
is obtained with the block-spin or ``coarse-graining" analysis: An
average over
a group of spins within a characteristic ``hypercube'' of dimension
$(\Lambda/s)^{-d}$ greater than the original elementary lattice step
$\Lambda^{-1}$,  defines the new average field.  The Hamiltonian
is built in terms of the new  ``blocked" average field  by
``integratingout"
the non-relevant degrees of freedom between $\Lambda$ and
$\Lambda/s$.
As consequence,
this procedure generates all the effective interactions (see
\cite{ma} for an
introduction).
Then by a global rescaling of the fields to the previous value of the
cutoff,
the action undergoes a renormalization group transformation that
governs an
infinite set of couplings.

In field theory the renormalization group equations are derived from
the
scaling behavior of the
Green's functions. The relation between the bare and the renormalized
vertex
functions,
\be
\Gamma_{B}^{\left (n\right )}(x^{1},...,x^{n};
\lambda^{i}_{B},\Lambda)=Z^{-n/2}(\lambda^{i}_{B},\Lambda/\mu)
\Gamma^{
(n)}_{\sf
ren}(x^{1},...,x^{n};
\lambda^{i}_{\sf ren}, \mu)
\ee
where $\Lambda$ and $\mu$ are the cutoff and a low-energy scale
respectively,
can be read either in terms of the renormalized or the bare
couplings. In one
case, by imposing the invariance of the renormalized vertex functions
under the
action of the diffeomorphisms generated by the vector field
$\Lambda\partial/\partial\Lambda$
in the parameter space, at fixed $\lambda^{i}_{\sf ren}$ and $\mu$,
one
obtains the renormalization group equation in the bare scheme.
Conversely, the
nondependence of the bare vertices from the low energy scale $\mu$
at fixed
bare couplings expresses the renormalization group equations for the
renormalized quantities. The two schemes can be formulated both in
a momentum
as well as in acoordinate space representation, but the
renormalized scheme
has a more clear physical meaning because it follows the evolution of
physical,
renormalized quantities.

In a nonflat background the loss of Poincar\'{e} invariance prevents
one from
defining a global momentum space. Thus, as suggested by Nelson and
Panangaden
\cite{np} the ultraviolet flow
of a generic interacting quantum field can be analyzed by looking at
the
scaling behavior of the field under a rescaling of the metric
$g_{\mu\nu}\rightarrow g_{\mu\nu}/s$. This can be intuitively
regarded as a
rescaling of the geodesic distance between two points, and it
is equivalent to introducing a local Riemann coordinate when the
points in the
argument of the Green's functions become close. The derivation
of the
renormalization group equations in a curved spacetime strictly
resembles the
procedure above outlined for the flat case. This is not surprising
because the
very-high-frequency modes will be increasingly insensitive to the
local
curvature as the cut-off tends to infinity. But by construction this
approach
can be reliable only to follow the evolution  of the UV relevant
coupling
constants.

In recent works the intimate connection between renormalization
group flow in the Kadanoff-Wilson formulation and the usual
concept
of
renormalizability in field theory has been deeply discussed (see
\cite{polchi})
and several efforts have been done to investigate the possibility of
formulating the block-spin approach in the framework of a quantum
field
theory.

The analogue of the Kadanoff transformation for the field variable
$\phi(x)$
is in principle constructed by means of the blocked field, the
average of the
original field in the characteristic volume $V_d$
\be
\phi_{V_d}=\frac{1}{V_d}\int_{V_d}{d^dx\phi(x)}.
\ee
One is interested in studying
the $V_d$ dependence of the distribution for the blocked field
\be
\eta(\Phi)=\langle
\delta(\Phi-\frac{1}{V_d}\int_{V_d}{d^dx\phi(x)})\rangle=\int D[\phi]
e^{-S[\phi]} \delta(\Phi-\phi_{V_d}).\label{ppa}
\ee
An equivalent realization can be carried out in a momentum-space by
defining
the coarse-grained field as
\be
\phi_k(x)=\int d^dy\rho_k(x-y)\phi(y)\label{cdo}
\ee
where the smearing function $\rho_{k}(x)$ acts as a projector on the
low-energy
region of momenta $q<k$ and in the configuration space rapidly decays
for
$1/k<|x-y|$. This latter alternative scheme turns out to be more
suited to a
perturbative analysis.  In fact, by using an O(4) smearing function
of the
kind (\ref{cdo}), Wetterich \cite{we} formulates  an ``average"
effective
action describing the fluctuations of the field average as the
``observational"
 region $1/k^4$ changes (see \cite{fk1} ). In a more recent work
Liao and
Polonyi \cite{lp,lp1} explicitly construct the Kadanoff transformation
and
calculate the renormalized action non perturbatively, with the
loop-expansion.
Their approach has the advantage of generating the renormalized
action fully,
with either  relevant and irrelevant terms as well, and clarifies in
what
respect the two formulations of the renormalization
group in statistical mechanics and field theory are equivalent.

A realization of the ``coarse-grained" effective action in curved
spacetime was
first given by Hu \cite{huh} to discuss the back reaction problem in
the
inflationary cosmological
models, and by Sinha and Hu \cite{sh} to examine the validity of the
minisuperspace approximation in quantum cosmology.

In this work a formulation of the Kadanoff-Wilson renormalization
group
approach in  curved space in the case of slowly varying background
manifold is
presented. The approach, borrowed
from \cite{we,lp}, represents an improvement of the perturbative
traditional
scheme.
Explicit expressions of the blocked action
are given, in the case of the Einstein universe, within the
loop-expansion
approximation at one loop  by means of the generalized \cite{hw}
$\zeta$-function
method.  When the expansion in the kinetic energy is possible, then a
local
potential $U_{n}(\Phi)$ governs fluctuations that have been averaged
in the
characteristic space volume $\Omega_{n}\sim(a/n)^3$
- here $a$ is the radius of the Universe.
It is found that the distribution of the block variables
\be
\eta_{n}(\Phi)=\langle\delta(\Phi-\phi_{n}(x))\rangle\approx
e^{-\Omega_{n}tU_{n}(\Phi)},
\ee
where $t$ is the time extent of the system, is strongly peaked around
two
non symmetric vacua as the ``observational'' space volume becomes
small enough.
This confirms the flat space result of \cite{lp} supporting the
conjecture that
the field distribution presents a domain structure even in the
symmetric phase.
We find that  in the strong curvature regime this effect is more
pronounced and
the crossover scale approaches the infrared as the curvature
increases.

The renormalization group analysis is employed by a differential
equation for
the local potential.
The behavior of the $\beta$ functions near the UV fixed point
reproduces the
perturbative flow for the relevant vertices. But the renormalization
group
approach based on the coarse-graining procedure traces the evolution
of all the
new interaction terms arising from the coupling of the field with
gravity as we
move towards the low energy region. In particular, the flow of the
conformal
coupling is analyzed in the improved scheme near the infrared end of
the UV
region. In particular, the infrared fixed point for a massive model
is present
for arbitrary renormalized coupling strength, and $\xi_R=1/6$ is not
an
infrared attractor.

The organization of the work is the following. In Sec. I the general
lines of
the blocking transformation in the case of static, spatially almost
homogeneous
spacetime are presented. In Sec. II the specific case of a closed
Einstein
Universe is treated in detail. In Sec. III the renormalization group
equation
is formulated and the infrared scaling around the UV fixed point is
discussed.
Sec. IV is devoted to the conclusions while  the relevant details of
the
computation are given in the Appendix.

\section{Blocking transformation}
Let the Euclidean section $\Omega$ of the spacetime
be a compact, boundaryless, Riemannian manifold.
The Euclidean bare action for the matter field reads:
\be
S[\phi(x)]=\int
d\Omega(x)\{\frac{1}{2}(\nabla_{\mu}\phi(x))^2+V(\phi(x))\}
\ee
where the bare potential is given by
\be
V(\phi(x))=\frac{1}{2}m_{B}^{2}\phi^{2}(x)+\frac{1}{2}\xi_{B}
R(x)\phi(x)^2+\frac{1}{4!}\lambda_{B}\phi^{4}(x)
\ee
$d\Omega(x)=g^{1/2}d^{4}x$ is the invariant four-volume  $g$, is the
determinant
of the Euclidean metric on the manifold, $\phi$ is a real scalar
field and
$R$ is the Ricci scalar. The Laplace-Beltrami operator  acting on
scalars is
\be
\nabla_{\mu}\nabla^{\mu}=\frac{1}{\sqrt{g}}\partial_{\mu}(\sqrt{g}
g^{\mu\nu}\partial_{\nu}).\label{oper}
\ee
The spectrum of this operator is discrete; the eigenfunctions
${\psi_{n}(x)}$
form a complete set in the space of the square integrable functions
$L^2(\Omega)$, and satisfy
\be
\nabla_{\mu}\nabla^{\mu}\psi_{n}(x)=-\omega_n^2\psi_{n}(x),
\ee
with the orthonormality condition
\be
\int_{\Omega}d\Omega(x)\psi_{n}(x)\psi^{*}_{m}(x)=\delta_{m n}
\ee
where the astrerisk means complex conjugation. A generic element in
$L^2(\Omega)$
reads
\be
\phi(x)=\sum_{n=0}^{+\infty}c_{n}[\phi]\psi_{n}(x),\label{expafi}
\ee
where in this concise notation the summation is understood even in
subspaces
spanned by degenerate eigenfunctions (if present). $n=0$ corresponds
to the
homogeneous mode, constant function on $\Omega$.

We define a new field, analogous to the block-spin variable
introduced in
statistical mechanics, $\phi'(x)=\phi_{n}(x)$, in the following way:
\be
\phi_{n}(x)=\int d\Omega(x')\rho_{n}(x,x')\phi(x'),\label{smefi}
\ee
where
\be
\rho_{n}(x,
x')=\sum_{m=0}^{m=n}\psi_{m}(x)\psi^{*}_{m}(x')\label{sme}
\ee
is a generalized smearing function that introduces a sharp cutoff
on the
short distance modes.  In general we could have chosen any smooth
test function
that, compatibly with the generic group of symmetries of our
spacetime,  is
almost constant within a given volume $\Omega_{n}$ where the field is
averaged,
and rapidly decays outside that region.  It should be stressed that
different
realizations of the Kadanoff transformation could in principle be
achieved by
means of different smearing functions but the scaling properties near
the fixed
points will be independent of how the modes have been eliminated. We
note that
$\phi_{n}(x)\rightarrow\phi(x)$ when $n\rightarrow\infty$ and,
conversely,
$\phi_{n}(x)\rightarrow\phi_{0}$ when  $n\rightarrow 0$, being
$\phi_{0}$ the
homogeneous mode. Moreover, under integration
$\rho_{n}(x,x')=\rho_{n}(x',x)$.

The average action is therefore defined as:
\be
e^{-S'_{n}[\phi'(x),]}=\int
D[\phi]\prod_{x}\delta(\phi_{n}(x)-\phi'(x))e^{-S[\phi(x)]}
\label{norma}
\ee
in this way the field $\phi$ is constrained to have an average
$\phi_{n}(x)$
equal to a given configuration of the block variable $\phi'(x)$.
This
amounts to performing the path integration in Eq. (\ref{norma}) on the
``fast'' variables, the $m\geq n+1$ modes.
We define the functional generator of the Green's functions for the
blocked field:
\be
Z[J]=\int D[\phi']e^{-S'_n[\phi'(x)]+\int d\Omega J(x)\phi'(x)}=\int
D[\phi]e^{-S[\phi(x)]+\int d\Omega(x)J(x)\phi_{n}(x)}
\ee
therefore the partition function $Z$ is invariant under the blocking
procedure.
This property implies that
\be
e^{-S'_{n}[\phi'(x)]}\nonumber
\ee
represents the relative probability distribution for the blocked
field
$\phi'(x)$, the average of the original field $\phi(x)$ to the
characteristic
volume $\sim 1/\omega^4_n$, associated with the n-eigenmode
considered.  The
mean value of a generic field operator for the blocked system is
related with
the previous one in the following manner:
 \bea
&&\langle O[\phi']\rangle
=Z^{-1}\int D[\phi']O[\phi']e^{-S'[\phi']}\nonumber\\[2mm]
&&=Z^{-1}\int
D[\phi']D[\phi]O[\phi']\prod_{x}\delta(\phi_{n}(x)-
\phi'(x))e^{-S[\phi]}\nonumber\\[2mm]
&&=Z^{-1}\int D[\phi]O[\phi_{n}]e^{-S[\phi]}=\langle
O[\phi_{n}]\rangle
\eea
This  means that  $\langle O[\phi']\rangle=\langle O[\phi]\rangle$
for all the
operators that are functions of the zero-momentum component of the
field. Thus
the mean value of the field is preserved by the blocking
transformation:
$\langle\phi'\rangle=\langle\phi\rangle$. Similarly we see that for
the
$p$-point Green's functions,
\be
\langle\phi'(x_{1})...\phi'(x_{p})\rangle\sim\langle
\phi(x_{1})...\phi(x_{p})\rangle
\ee
if $n$ is sufficiently high.

The blocking procedure generates in (\ref{norma}) all the effective
interactions corresponding to the low-energy
$m\leq n$ in a generic static spacetime.   However, in order to
perform
explicit calculations we consider the case of  almost-homogeneous
space
sections: If the disomogeneities of the background manifold are
``smooth''
within the characteristic regions in which the original field  has
been
average out (one can always think to the limiting case of homogeneous
space
sections) then it is possible to carry out a local derivative
expansion around
any given point on the manifold (see\cite{hu1}):
\be
\phi'^2(x)=\phi'^2(x')+\nabla_{\mu}\phi'^2(x')(x-x')^{\mu}+
\frac{1}{2}\nabla_{\mu}\nabla_{\nu}\phi'^2(x')(x-x')^
{\mu}(x-x')^{\nu}+...\label{expa}
\ee
Therefore we write, according to the Landau-Ginzburg lesson, the
following
functional form for the new free-energy:
\be
S'_{n}[\phi']=\int
d\Omega(x)\sum^\infty_{i}U^{(i)}_{n}(\nabla_{\mu}\phi'(x))
\label{general}
\ee
where $U^{(i)}_{n}(\nabla_{\mu}\phi'(x),)$ is a general polynomial of
degree
$i$ in the covariant derivative $\nabla_{\mu}$. Note that the
gradient
expansion becomes more accurate as we scale in the infrared. To study
the
fluctuations around the $\phi'(x)=\phi_{n}(x)$ configuration, one
rewrites the
operator projector kernel in (\ref{norma}) as
\be
\prod_{x}\delta(\phi_{n}(x)-\phi'(x))=\exp {-M^{2}\int
d\Omega(x)(\phi_{n}(x)-\phi'(x))^2},
\ee
where $M$ is a massive constant that in a more refined analysis could
be
mode dependent. In this case it represents a measure for the mean
deviation  from the $\phi_{n}(x)=\phi'(x)$ configuration and it is
taken as being far greater than all the relevant masses in the
theory at the end of the computation. The constrained action in
(\ref{norma})
reads
\be
e^{-S'_{n}[\phi'(x)]}=\int D[\phi]e^{-S_{n}[\phi(x),\phi'(x)]}
\label{con}
\ee
where
\bea
&&S_{n}[\phi,\phi']=\int
d\Omega(x)\{\frac{1}{2}(\nabla_{\mu}\phi(x))^2+
\frac{1}{2}m_{B}^2\phi^2(x)+\frac{1}{2}\xi_{B}
R\phi^2(x)+\frac{1}{4!}\lambda_{B}\phi^4(x)+\nonumber\\[2mm]
&&+M^2(\phi_{n}(x)-\phi'^2(x)).\label{con1}\}
\eea
To identify the non derivative terms generated
in the renormalized action it is enough to evaluate (\ref{general})
for a
constant field configuration
$\phi'(x)=\Phi$
so that the renormalized action has the functional form:
\be
S'_{n}[\Phi]= \Omega U_{n}(\Phi) \quad\quad\quad
U_{n}(\Phi)=\frac{1}{2}\mu^2(n)\Phi^2+\frac{\lambda(n)}{4!}\Phi^4
+\dots
\label{locapo}
\ee
where $U_{n}(\Phi)$ is the average potential.  Equation (\ref{con1})
becomes
 \bea
&&S_{n}[\phi,\Phi]=\int
d\Omega(x)\{\frac{1}{2}(\nabla_{\mu}\phi(x))^2+
\frac{1}{2}m_{B}^2\phi^2(x)+\frac{1}{2}\xi_{B}
R\phi^2(x)+\frac{1}{4!}\lambda_{B}\phi^4(x)+\nonumber\\[2mm]
&&+M^2\Phi^2-2M^2\phi(x)\Phi\}+M^2\int d\Omega(x)
d\Omega(x')\rho_{n}(x,x')\phi(x)\phi(x')\label{con2}.
\eea

We evaluate the renormalized action in the loop expansion.
In choosing the saddle point in (\ref{con})
we have to locate the minimum of (\ref{con2}),
obtaining the non local motion equation
\bea
&&0=\frac{\delta S_{n}[\phi,\Phi]}{\delta
\phi(x)}=-\nabla_{\mu}\nabla^{\mu}\phi(x)+m_{B}^{2}\phi(x)+\xi_{B}
R\phi(x)+\frac{\lambda_{B}}{3!}\phi^3(x)\nonumber\\[2 mm]
&&+2M^2\int d\Omega(x')\rho_{n}(x',x)\phi(x')-2M^2\Phi.
\label{classic}
\eea
Consistently with the hypothesis of  weak disomogeneities, we assume
that in
this approximation $[\nabla_{\mu}\nabla^{\mu}, R(x)]\sim 0$, so that
Eq. (\ref{classic}) for a generic constant field configuration
$\phi(x)=\phi_{0}$ becomes
\be
\left ( 2M^2+m^2_{B}+\xi_{B}R+\frac{1}{3!}\lambda_{B}\phi_{0}^2
\right )
\phi_{0}=2M^2\Phi .\label{moto}
\ee
By substituting in (\ref{con2}) one finds
\be
U_{n}(\Phi)=-\frac{\lambda_{B}}{4!}\phi_{0}+
M^2\Phi(\Phi-\phi_{0}).\label{solumo}
\ee
Therefore at the tree level in this case the effect of
``coarse graining''
is not present and we recover the classical potential
\be
U_{n}(\Phi)=\frac{1}{2}m_{B}^2\Phi^2+\frac{1}{2}\xi_{B}
R\Phi^2+\frac{\lambda_{B}}{4!}\Phi^4+O(1/M^2) \label{clapo}
\ee
To include the quantum corrections one writes
$\phi(x)=\phi_0+\chi(x)$
where $\chi(x)$ represents the quantum fluctuations. At
the one-loop order in the quantum corrections the
renormalized potential
is given by
\be
S'_{n}[\Phi,\chi,]=\Omega V(\Phi)+\Delta S_{n}[\Phi,\chi],
\ee
where
\be
\Delta S_{n}[\Phi,\chi]=\frac{1}{2}\int d\Omega(x) d\Omega(x')\left
.\frac{\delta^2
S_{n}}{\delta\phi(x)\delta\phi(x')}\right|_{\Phi}\chi(x)\chi(x')
\ee
with
\be
\left .\frac{\delta^2 S_{n}}{\delta\phi(x)\delta \phi(x')}\right
|_{\Phi}=(-\nabla_{\mu}\nabla^{\mu}+m_{B}^2+\xi_{B}
R+\frac{\lambda_{B}}{2}\Phi^2)\delta(x,x')+2M^2\rho_{n}(x,x').\label{f
unde}
\ee
The contribution of loops with internal quantum numbers
$m\leq n$ is  suppressed by a factor $1/M^2$; thus the
renormalized potential at one loop in the loop-expansion can be
rewritten
\bea
&&U_{n}(\Phi)=V(\Phi)+\frac{1}{2\Omega}\ln {\rm Det}\left
.\frac{\delta^2
S_{n}}{\delta\phi(x)\delta \phi(x')}\right |_{\Phi}\nonumber\\[2 mm]
&&=V(\Phi)+\frac{1}{2\Omega}{\sf
Tr}_{n+1}\ln[(-\nabla_{\mu}\nabla^{\mu}+m_{B}^2+\xi_{B}
R+\frac{\lambda_{B}}{2}\Phi^2)\delta(x,x')], \label{avepot}
\eea
where ${\sf Tr}_{n+1}$ means that in the trace operation only modes
with $m\geq
n+1$ are retained. We set $n+1\rightarrow n$ so that when $n=0$
expression
(\ref{avepot}) matches the familiar effective potential \cite{cw,fk1}
defined
usually by means of a Legendre transform of the Schwinger function.
We observe
that the spectrum of
$\delta^2S_{n}/\delta\phi(x)\delta\phi(x')|_{\Phi}$ is
always positive if $\xi>0$ and the Ricci scalar is non-negative. This
ensures
that the path integral is dominated by homogeneous field
configurations. In the
following we consider $\xi>0$.
\section{Einstein universe}
If we work initially at finite temperature, the Einstein universe has
topology
$S^1\times S^3$. The high level of symmetry of this spacetime allows
one to
perform explicit calculations. Let the Euclidean metric be
\be
ds^{2}_{E}=d\tau^2+a^2[d\chi^2+\sin^2\chi(d\theta^2+\sin^2\theta
d\varphi^2)]
\ee
where $\chi,\theta,\varphi$ label $S^3$, $a$ is the radius, and
$\tau$ spans
the euclidean time. In this case the eigenfunctions of the
Laplace-Beltrami
operator are known: they factorize in the product of plane waves
$e^{-2\pi i
n\tau/\beta}$
- here $\beta$ is the invariant time - and the hyperspherical
functions on
$S^3$:
\be
Y_{nlm}(\chi,\theta,\varphi)=N_{nlm}Y_{lm}(\theta,\varphi)
\sin^{l}\chi C^{l+1}_{n-l}(\cos\chi)
\ee
where $N_{lmn}$ is a normalization factor,
$C^{q}_{p}(x)$ represents a Gegenbauer polynomial, and is
$Y_{lm}(\theta,\varphi)$ the usual spherical harmonic on $S^2$. The
quantum
numbers $\{n,l,m\}$ are
$n=0,1,2,...;l=0,1,2,...,n;m=-l,-l+1,...,l-1,l;$. The
eigenvalues of the Laplace-Beltrami operator are found to be
$\omega^2(n,p)$=
$(2\pi p/\beta)^2 +\lambda^2_{nlm}$ where
$\lambda^2_{nlm}=a^{-2}((n+1)^2-1)+m_{B}^2+\xi_{B}
R+\frac{\lambda_{B}}{2}\Phi^2$ are the eigenvalues on the
spatial sections with degeneration $d=\sum_{l=0}^{n}(2l+1)=
(n+1)^2$. Here $a^2=6/R$ is the radius of $S^3$.

In this manifold a possible realization of  the ``coarse graining''
procedure
can be achieved by integrating out the fast fluctuating space modes
on $S^3$
and by leaving unconstrained the time evolution of the system.  This
case is
important because it allows us to study the spatial distribution of the
fluctuations when the observational characteristic volume shrinks.
To this aim we use the
following smearing function  (see \cite{lp1} for an analogous $O(3)$
invariant
coarse-graining in flat space)
\be
\rho_{n}(P,P')=\sum_{n'=0}^{n}\sum_{l=0}^{n'}
\sum_{m=-l}^{m=l}Y_{nlm}(P)Y^{*}_{nlm}(P')\label{medi}
\ee
and $P,P'\in S^3$.
The functional determinant in expression (\ref{avepot}) is calculated
by means
of the generalized zeta-function $\zeta(s)$:
\be
U_{n}(\Phi)=V(\Phi)+\frac{1}{2\Omega}(-\zeta'(s)|_{s=0}) \label{zeta}
\ee
where $\Omega=2\pi^2\beta a^3$ is the total volume. In our case, in
the low-temperature limit we have
\be
\zeta(s)=\frac{\beta\mu^{2s}}{2\pi} \int
dp\sum_{n'=0}^{\infty}\sum_{l=0}^{n'}\sum_{m=-l}^{m=l}
(\omega^2(p,n)+4M^2{\pi}\delta(p)\Theta_{n,n'})^{-s} \label{zeta1}
\ee
with $\Theta_{n,m}=1$ for  $m\leq n$ and zero elsewhere.
Explicit evaluation of expression (\ref{zeta1}) produces a
regularized
expression for the
local potential that is dependent  on the undetermined constant $\mu$
(see the
appendix for the relevant details). One finds
\be
U_{n}(\Phi)=V(\Phi)-\frac{1}{64\pi^2 a^4}\lbrace
(\tilde{V}''(\Phi)-1)^2
[\ln \frac{\mu^2}{R/6}+\frac{3}{2}]+ \Upsilon_{n}(\Phi)\rbrace
\label{reno}
\ee
where now the range of  $n$ starts at $n=1$.  In the above formulas
 $\tilde{V}''(\Phi)=a^2V''(\Phi)=a^2(m_{B}^2+\xi_{B}R+\lambda_{B}\Phi^
2/2)$ and
$\Upsilon_{n}(\Phi)$ is a function of $\Phi$ and $R$ whose explicit
form is
given in the Appendix and whose low curvature limit
$\tilde{V}''(\Phi)\gg 1$ ,
for $n=1$ is
\be
\frac{R^2}{144\pi^2}\Upsilon_{1}(\Phi)
\sim\frac{1}{64\pi^2}(V''(\Phi)-R/6)\ln\frac{V''(\Phi)-R/6}{R/6}
\label{aego}
\ee
A subtraction scale is chosen in order to eliminate the $\mu$
dependence. One
writes  $m_{B}^2=m_{R}^2+\delta m^2$, $\xi_{B}=\xi_{R}+\delta\xi$,
$\lambda_{B}=\lambda_{R}+\delta\lambda$ in the bare expression
(\ref{reno}),
and introduces the following renormalization conditions for the
counterterms:
\bea
&&m_{R}^2=\left
.{\frac{\partial^2U_{n=1}}{\partial\Phi^2}}
\right|_{\Phi=R=0}\hspace{1cm}\xi_{R}=
\left .{\frac{\partial^3U_{n=1}}{\partial\Phi^2\partial
R}}\right|_{\Phi=R=0}\hspace{1cm}\lambda_{R}=\left
.{\frac{\partial^4U_{n=1}}{\partial\Phi^4}}\right|_{\Phi=R=0}
\label{point}
\eea
\vspace{0.3 cm}
where for the sake of simplicity the renormalization point  has been
chosen in
$R=0$.  From expression (\ref{aego}) and from Eq. (\ref{point}) (see
Appendix )
we have
\bea
&&U_{n}(\Phi)=\frac{1}{2}m_{R}^2\left [1+\frac{\xi
R}{m_{R}^2}-\frac{\lambda_{R}}{64\pi^2}\left
(1+\frac{3R(\xi_{R}-1/6)}{m_{R}^2}\right )\right
]\Phi^2+\frac{\lambda_{R}}{4!}\left
(1-\frac{9\lambda_{R}}{64\pi^2}\right
)\Phi^4+\frac{R^2{\cal{I}}_{n}(\Phi)}{144\pi^2}\nonumber\\[2mm]
&&+\frac{1}{32\pi^2}\left (m_{R}^2+\left (\xi_{R}-\frac{1}{6}\right
)R+\frac{1}{2}\lambda_{R}\Phi^2\right
)^2\ln\frac{n\sqrt{R/6}+\sqrt{m^2_{R}+\left (6\xi_{R}+n^2-1\right
)R/6+\lambda_{R}\Phi^2/2}}{m_{R}}\nonumber\\[2mm]
&&-\frac{3n\sqrt{6R}}{576\pi^2}\sqrt{m^2_{R}+\left
(6\xi_{R}+n^2-1\right
)\frac{R}{6}+\frac{\lambda_{R}\Phi^2}{2}}\left (m^2_{R}+\left
(6\xi_{R}+2n^2-4n-1\right )\frac{R}{6}+\frac{\lambda_{R}\Phi^2}{2}
\right )
 \label{reno1}
\eea
where the following counterterms have been introduced
\bea
\delta
&&m^2=-\frac{\lambda_{R}}{32\pi^2}m_{R}^2(\ln\frac{m_{R}^2}
{\mu^2}-1),\quad
\delta\xi=-\frac{\lambda_{R}}{32\pi^2}(\xi_{R}-\frac{1}{6})
\ln\frac{m_{R}^2}{\mu^2}\nonumber \\[2 mm]
&&\quad\quad\quad\quad\quad\quad\quad
\delta\lambda=-\frac{3\lambda_{R}^2}{32\pi^2}\ln\frac{m_{R}^2}{\mu^2}
\eea
and ${\cal{I}}_{n}(\Phi)$ is an integral function of  the field  and
the
renormalized parameters - see Appendix.
We note that  for $R=0$ expression $(\ref{reno1})$ reproduces the
standard
Coleman-Weinberg result $\cite{cw}$.  As shown in the appendix,
the local
potential for $n=1$ matches the form of  the effective potential in
the Einstein universe: there is no spontaneous symmetry breaking
induced by
curvature if $\xi_{R}>0$  (note that expression (\ref{reno1}) is
always real),
even though the quantum contribution tends to drive the system in the
spontaneously broken phase \cite{hu}.  However as $n$ increases the
new
action governs the distribution of the
field whose average has been performed
within the characteristic space volume $\Omega_{n}\sim (a/n)^3$. In
general
the location of the minimum
in the blocked action
depends on the observation domain $\Omega_{n}$. It is natural to
expect  that when the average is performed over regions
far larger than the characteristic regions
in which the fluctuations of the field are correlated, the details of
the original distribution are not yet detectable and statistical
equilibrium of the system is reached in $\Phi=0$: This is encoded in
the convex
character of $U_{n}(\Phi)$ for low $n$. But
when the ``observational''
region is small enough, the structure of the original
field variables may show up. In fact a
numerical investigation of (\ref{reno1}) shows that after some
critical value
of $n=n_{cr}$, the blocked potential
is no longer convex, and non trivial minima appears - see Fig.1 and
Fig.2.
Below that
scale the convexity of the potential is recovered.
In the case where the curvature is not strong, one can gain
more insight into this phenomenon  by introducing a local momentum
space around a generic point in the manifold and  examining the
Minakshisundaram-Pleijel-DeWitt expansion\cite{mi,de} to first order.
This is
possible because the presence
of non trivial minima occurs for high values of $n$, for small $R$,
and
therefore we expect the
local-momentum approximation to hold (for example one finds $n_{\sf
cr}\sim
10^3$ for $R/m^2_{R}\sim 10^{-2}$). In this case is convenient to use
the translation-invariant smearing function
\be
\rho_k(x)=\int_{|p|<k}\frac{d^4p}{(2\pi)^4}e^{-ipx}
\ee
to define the blocked field (\ref{smefi}), and the field is averaged
in a region
of  extent $\sim 1/k^4$. By using a cut-off regularization procedure,
the
one-loop blocked potential is given by
\be
U_{k}(\Phi)=V(\Phi)+\frac{1}{64\pi^2}\int_{k}^{\Lambda}
dpp^3{\rm ln}\left (1+
\frac{V''(\Phi)-R/6}{p^2}\right ) \label{varano}
\ee
and Eqs. (\ref{point}) give the following counterterms
\bea
&&\delta m^2=-\frac{\lambda_{R}}{32\pi^2}\left (\Lambda^2+m_{R}^2\ln
\frac{m_{R}^2}{\Lambda^2}\right )\hspace{2
cm}\delta\xi=-\frac{\lambda_{R}}{32\pi^2}\left
(\xi_{R}-\frac{1}{6}\right )
\left (1+\ln\frac{m_{R}^2}{\Lambda^2}\right )\nonumber\\[2mm]
&&\quad\quad\quad\quad\quad\quad\quad\quad\quad\quad
\delta\lambda=-\frac{3\lambda_{R}^2}{32\pi^2}
\left (1+\ln\frac{m_{R}^2}{\Lambda^2}\right )
\eea
The average potential in this
approximation reads
\bea
&&U_{k}(\Phi)=\frac{1}{2}m_{R}^2
\left [1+\frac{\xi_{R}R}{m_{R}^2}-\frac{\lambda_{R}}
{64\pi^2}\left (1+\frac{k^2+3R(\xi_{R}-1/6)}{m_{R}^2}\right )\right
]\Phi^2
+\frac{1}{4!}\lambda_{R}\left (1-\frac{9\lambda_{R}}
{64\pi^2}\right )\Phi^4\nonumber\\[2mm]
&&+\frac{1}{64\pi^2}\left [ \left (m_{R}^2+\frac
{1}{2}\lambda_{R}\Phi^2+\left (\xi_{R}-\frac{1}{6}\right )R\right
)^2-k^4\right
]
\ln\frac{k^2+m_{R}^2+\frac{1}{2}\lambda_{R}\Phi^2
+\left (\xi_{R}-\frac{1}{6}\right )R}{m_{R}^2}\label{door}
\eea
Let us consider the dimensionless running parameter
\be
\frac{1}{m^2_{R}}\left .\frac{\partial^2U_{k}(\Phi)}{\partial\Phi^2}
\right|_{\Phi=0}={\cal M}^2(k)
\ee
In the limit of $k^2\gg m_{R}^2+\xi_{R}R$
one finds
\be
{\cal
M}^2(k)=1+\frac{\xi_{R}R}{m_{R}^2}-\frac{\lambda_{R}k^2}{32\pi^2}
+\frac{\lambda_{R}}{32\pi^2}\left [1+\left (\xi-\frac{1}{6}\right
)\frac{R}{m_{R}^2}\right ]\ln\frac{k^2}{m_{R}^2}\label{cr}
\ee
It is interesting to observe the competition between the two scales
$k$ and
$R^{1/2}$. When the logarithmic term is dominant  in (\ref{cr})
${\cal M}^2(k)$
is  positive for $\xi_{R}>0$, and the only minimum
for (\ref{door})  is at $\Phi=0$. But as the observational scale
shrinks, the
linear term in $k^2/m_{R}^2$ governs, and the local potential
presents two
degenerate non symmetric minima at some $k=k_{cr}$. Conversely, if
the matter
field is strongly coupled with gravity at classical level $\xi_{R}{\
\lower-1.2pt\vbox{\hbox{\rlap{$>$}\lower5pt\vbox{\hbox{$\sim$}}}}\
}1$,
the linear term in $\xi_{R}R/m_{R}^2$ may eventually dominate, and
therefore
for some $R=R_{\sf cr}$ it will turn the maximum in $\Phi=0$ in to a
minimum.
The value of $R_{\sf cr}$ in this approximation can be read from
${\cal
M}(k)=0$
\be
R_{\sf cr}(k)=\frac{-m^2_{R}+\frac{\lambda_{R}}{32\pi^2}\left
(k^2-m_{R}
^2\ln\frac{k^2}{m_{R}^2}\right )}{\xi_{R}-\frac{\lambda_{R}}{32\pi^2}
\left
(\xi_{R}-\frac{1}{6}\right)\ln\frac{k^2}{m_{R}^2}}\label{cricri}
\ee
At the leading order, for a conformally coupled field one finds
\be
R_{\sf cr}(k)\simeq
\frac{3\lambda_{R}k^2}{16\pi^2}-m^2_{R} \label{cri1}
\ee
The expression (\ref{cricri}) is independent of the
background spacetime but it is reliable only in the small curvature
approximation $R\ll k^2$ where the local-momentum analysis works if
the
coupling strength $\lambda_{R}$ is weak enough - see relation
(\ref{cri1}). The
new minima in this
approximation are located at
\be
\Phi^2=m_{R}^2\frac{-1-\frac{\xi_{R}R}{m^2_{R}}+
\frac{\lambda_{R}}{32\pi^2}\left (
\frac{k^2}{m_{R}^2}-\left (1+\left (\xi_{R}-1/6\right
)\frac{R}{m_{R}^2}\right
)\ln\frac{k^2}{m_{R}^2}\right )}{\frac{\lambda_{R}}{6}\left
(1-\frac{3\lambda_{R}}{32\pi^2}\right )+\frac{\lambda_{R}^2}{64\pi^2}
\ln\frac{k^2}{m_{R}^2}}
\ee
The presence of degenerate minima in the local potential
at a particular value of $n$ and $R$, is a different phenomenon from
the
spontaneous symmetry breaking, peculiar to the lowest energy mode.
As clarified by \cite{lp}, it can be argued that the field presents a
domain
structure even in the purely symmetric phase ${\langle \phi
\rangle}_{vac}=0$.
The path integral is dominated by domains of characteristic size
$\iota$ where
$\phi(x)\sim\pm\bar{\Phi}$.  Then, if we average the field over
regions of
linear extension $k^{-1}\gg \iota$ the fluctuations between the
domains become
uncorrelated and the distribution of the blocked variables is
centered around
$\Phi \sim 0$.
But as the observational scale is comparable with the size $ k_{\sf
cr}^{-1}\sim \iota$ of the domains, the fluctuations are governed by
a
distribution that is strongly peaked around two non-symmetric
degenerate
values $\pm\bar{\Phi}$ of the field. In other words, as  we move
from the
ultraviolet towards the infrared domain, a crossover between a
double peak and
a symmetric distribution occurs.  When we average in larger regions
the
Gaussian limit is approached, according to the central limit theorem,
and,
since the correlations among domains
typically decay as $\sim e^{-m/k}$ where $m$, the mass gap, is the
inverse of
the correlation length, one concludes that $k_{\sf cr}\sim m$.

In curved spacetime
the other relevant length is the local
curvature $R$ that couples quadratically with the field at the
classical level,
and enters in the propagator as a pure geometric contribution as well.
If  $R\gg
m^2_{R}$, the local-momentum approximation is questionable.  Unless
exact
expressions for the propagator are available,  one has to resort to
alternative
approaches, by taking into account the global topology of the
spacetime, to
describing the running of the ``effective'' correlation length
$1/{\cal
M}(n)m_{R}$ in the strong curvature regime.  In the case of the
Einstein  universe
the large $R$ and large $n$ approximation give, to the
leading order,
\be
{\cal M}^2(n)=1+\frac{R}{m_{R}^2}\left
(\xi_{R}-\frac{\lambda_{R}n(n-1)}{96\pi^2}\right )=0\label{intr}
\ee
for the critical scale.
The crossover region is therefore shifted toward the infrared domain
by an
increase of curvature as the coefficient of  $R/m^2_{R}$ becomes
negative - see
Figs.1 and 2.  This result is interesting, we believe,  because it
illustrates how the geometry of the manifold determines the interplay
between
the two relevant lengths $m_{R}$ and $R$ in setting the crossover
scale. Note
that $n$ in this case cannot be approximated by a continuous variable.
For
instance, $\lambda_{R}=0.1$ and $R/m_{R}^2\sim 1$, eq.(\ref{intr})
yields
$n_{\sf cr}=138$.

\section{renormalization group analysis}

The invariance of the partition function $Z[J]$ under coarse-graining
yields
the following equation for the local potential
\be
U_{n}(\Phi')-U_{n-\Delta n}(\Phi')=\Delta n\Gamma_{n}[U_{n}(\Phi')]
\label{func}
\ee
where $\Gamma[U_{n}]$ represents the contribution of the eliminated
modes from
the original action in the interval $(n, n-\Delta n)$. This ensures
\cite{polchi} that the generator of the Green's functions $Z[J]$ and
its
functional derivatives are unchanged when the modes are eliminated in
the
starting action. Equation (\ref{func}) governs the scaling of all the new
interaction
terms generated by the blocking procedure to compensate for the
eliminated
modes.
In a nonflat background the explicit form of  $\Gamma_{n}$ is
generally
dependent on the geometry of the  manifold as we approach the deep
infrared
region. Since, for a compact manifold, the spectrum of the
Laplace-Beltrami
operator is discrete, Eq. (\ref{func}) becomes a
partial finite-difference
equation for the local potential whose solution close to the low $n$,
region
determines in principle the relevant operators at the infrared fixed
point. At
one loop in the loop expansion we see that the contribution of the
eliminated
modes in the ``momentum shell''' $(n,n-\Delta n)$ is
\be
\Gamma_{n}[U_{n}(\Phi)]=\frac{1}
{2\Omega}\ln[\omega^2_{n}+\partial_{\Phi}^2U_{n}(\Phi)] \label{gam}
\ee
Although a solution of (\ref{func}) would in principle tell us how
the scaling
properties of the theory change as we approach the infrared, it
should be
stressed that the contribution of (\ref{gam}) may not be the leading
one in
determining the effect of the integrated modes on the evolution of
the
coupling constants .  In general higher order contributions in the
loop
expansions have to be retained as well.  It is however important to
study the
infrared scaling in the UV region to show  in what respect the
standard
realization of the renormalization group, based on the analysis of the scaling
properties of the
Green's function under a geodesic interval \cite{np}, and this
formulation,
obtained by the blocking, are equivalent.
To this aim we can work in the local-momentum approximation and in
this limit
(\ref{func}) reduces to
\be
k\frac{dU_{k}}{dk}=-\frac{k^4}{16\pi^2}\ln\left
(1+\frac{\partial^2_{\Phi}U_{k}-
R/6}{k^2}\right )\label{group1}
\ee
The blocking procedure generates new effective
vertices arising from the coupling of the matter field  with gravity.
Let us
write the  potential term in the general form
\be
U_{k}(\Phi,R)=\sum_{n=2}^{\infty}
\frac{\lambda_{n}+\xi_{n}R}{n!}\Phi^{n}\label{gene},
\ee
and introduce, for the following analysis, the dimensionless
variables
\be
R=\tilde{R}k^{2},\quad\Phi=\tilde{\Phi}k,\quad\lambda_{n}=
\tilde{\lambda}_{n}k^{4-n}\quad\xi_{n}=\tilde{\xi}_{n}k^{2-n}
\ee
The running coupling constants are defined
\be
\xi_n(k)=\left
.\frac{\partial^{n+1}U_{k}(\Phi,R)}{\partial\Phi^n\partial R
}\right |_{\Phi=R=0}\quad\quad\quad\lambda_n(k)=\left.
\frac{\partial^nU_{k}(\Phi,R)}{\partial\Phi^n}\right
|_{\Phi=R=0}\label{run}
\ee
Equation (\ref{group1}) in terms of  those variables reads:
\be
\lbrack k\frac{\partial}{\partial
k}-\tilde{\Phi}\frac{\partial}{\partial\tilde{\Phi}}-2\tilde{R}
\frac{\partial}{\partial\tilde{R}}+4\rbrack\tilde{U}_{k}(\tilde{\Phi},
\tilde{R})=-\frac{1}{16\pi^2}\ln[1+\partial^2_{\tilde{\Phi}}\tilde{U}_{k}
(\tilde{\Phi},\tilde{R})-\tilde{R}/6]\label{group2}
\ee
where $\Gamma=k^4\tilde{\Gamma}$ and $U=k^4\tilde{U}$. We saw that at
the tree
level
$\Gamma=0$ therefore (\ref{group2}) gives the solution
$\tilde\lambda_{n}$=$k^{n-4}C_{n}$, $\tilde{\xi}_{n}=k^{n-2}C'_{n}$,
thus for $n>4$ all the $\tilde{\lambda}_{n}$, and for $n>2$ all the
$\tilde{\xi}_n$ couplings are irrelevant. The above procedure could
be
generalized to describe a potential containing  more generic
interaction terms
of the form $\chi_{n,m}R^{n}\Phi^{m}$, however at the tree level a
simple
dimensional analysis suffices to classify the interactions.
In the deep UV region is $k^2\gg \partial^2_{\Phi}U_{k}(\Phi,R)-R/6$
and
(\ref{group2}) reads in this approximation
\bea
&&\lbrack k\frac{\partial}{\partial
k}-\tilde{\Phi}\frac{\partial}{\partial\tilde{\Phi}}-
2\tilde{R}\frac{\partial}{\partial\tilde{R}}+
4\rbrack\tilde{U}_{k}(\tilde{\Phi},\tilde{R})=-\frac{1}{16\pi^2}
[\partial^2_{\tilde{\Phi}}\tilde{U}_{k}(\tilde{\Phi},\tilde{R})-
\tilde{R}/6-\nonumber\\[2mm]
&&-\frac{1}{2}(\partial^2_{\tilde{\Phi}}\tilde{U}_{k}
(\tilde{\Phi},\tilde{R})-\tilde{R}/6)^2+...]
\label{appo}
\eea
{}From this expression one derives the evolution equations for all
the coupling
constants. This can be done by inserting (\ref{gene}) in (\ref{appo})
and by
factoring the coefficient of the generic power of the field. If we
arrest the
expansion at the relevant and marginal couplings,

\bea
 &&(k\frac{\partial}{\partial k}
+2)\tilde{\lambda}_{2}=-\frac{1}{16\pi^2}
\tilde{\lambda_{4}}(1-\tilde{\lambda}_2)+...
\nonumber\\[2 mm]
 &&k\frac{\partial\tilde{\xi}_{2}}{\partial
k}=\frac{1}{16\pi^2}\tilde{\lambda}_{4}(\tilde{\xi}_{2}
-\frac{1}{6})+...\nonumber\\[2mm]
&&k\frac{\partial\tilde{\lambda}_{4}}{\partial
k}=\frac{3}{16\pi^2}\tilde{\lambda}_{4}^2+...
\label{appo1}
\eea
where the ellipses stand for irrelevant couplings.
The system (\ref{appo1}) reproduces the standard perturbative
behavior for
interacting scalar field in curved spacetime \cite{np} for the
relevant
couplings, but this system handles the mixing of  all the new local
interactions along the renormalized trajectories as we
approach the infrared. This result can be
extended in the improved scheme where the interaction between modes
is taken into account by retaining the effect of the previously
integrated
modes when the cutoff is lowered. In this way one considers
the impact of the UV irrelevant operators during further
renormalization group transformations. Let us define the
renormalization group
coefficient functions (at a fixed value of the cutoff and the bare
parameters):
\bea
&&\frac{k}{m^2(k)}\frac{dm^2(k)}{dk}=
\gamma_{m}(\frac{m(k)}{k},\lambda(k),\xi(k))\nonumber\\[2mm]
&&k\frac{d\xi(k)}{dk}=\gamma_{\xi}(\frac{m(k)}{k},
\lambda(k),\xi(k))\nonumber\\[2mm]
&&k\frac{d\lambda(k)}{dk}=\beta(\frac{m(k)}{k},\lambda(k),\xi(k))
\eea
{}From (\ref{door}) and (\ref{run}) we find
\bea
&&\gamma_{m}(\frac{m_R}{k},\lambda_R,\xi_R)=
-\frac{\lambda_R}{16\pi^2}\frac{k^4}{m^2_{R}(m^2_{R}+k^2)}
\nonumber\\[2mm]
&&\gamma_{\xi}(\frac{m_R}{k},\lambda_R,\xi_R)=
\frac{\lambda_R(\xi_R-\frac{1}{6})}{16\pi^2}\frac{k^4}
{(k^2+m^2_{R})^2}\nonumber\\[2mm]
&&\beta_{\lambda}(\frac{m_R}{k},\lambda_R,\xi_R)
=\frac{3\lambda^2_{R}}
{16\pi^2}\frac{k^4}{(k^2+m^2_{R})^2}\label{scle}
\eea
Therefore the renormalization group improved  scale dependence of the
relevant
coupling constants  is given by the equations
\bea
&&k\frac{\partial m^2(k)}{\partial
k}=-\frac{\lambda(k)}{16\pi^2}\frac{k^4}{m^2(k)+k^2}\nonumber\\[2mm]
&&k\frac{\partial \xi(k)}{\partial
k}=\frac{\lambda(k)(\xi(k)-\frac{1}{6})}
{16\pi^2}\frac{k^4}{(k^2+m^2(k))^2}\nonumber\\[2mm]
&&k\frac{\partial \lambda(k)}{\partial
k}=\frac{3\lambda^2(k)}{16\pi^2}\frac{k^4}
{(k^2+m^2(k))^2}\label{infras}
\eea
with the initial conditions $m(0)=m_R, \xi(0)=\xi_R,
\lambda(0)=\lambda_R$.
It is important to stress that the conformal
coupling does not necessarily flow to the conformal value $\xi_R=1/6$
for
$k=0$.
In fact this ultimate set of equation shows the role played by the
mass gap in
determining the crossover region. Below the mass scale all the
fluctuations are
damped and the IR fixed point is present at $k=0$ for arbitrary values
of the
renormalized coupling constants $\lambda_R$ and $\xi_R$. However, in
the usual
regularization perturbative scheme
combined with the minimal subtraction prescription, one introduces
counterterms that are mass independent and this fixed point is
normally
missing for non vanishing coupling strength. But this
approximation is accurate only to study the scaling
in the deep UV region where all the relevant masses can be neglected.
\section{summary}
A realization of the coarse-grained
procedure has been presented  for a generally coupled scalar field in
the
Einstein universe, and the non derivative part of the renormalized
action has
been calculated at the one-loop approximation in the loop expansion. A
local
renormalized potential governs the spatial distribution of the
blocked
variables, the average of  the fluctuations in ``cubes'' of  size
$\Omega_{n}\sim
(a/n)^3$. This distribution is no longer centered on the symmetric
minimum
$\Phi=0$ as $\Omega_{n}\sim \iota^3$ where $\iota$ is the
characteristic linear
extent of   domains where the field assumes nonzero values. The
overall
geometry of the manifold determines  the crossover scale in the high
curvature
limit where the local methods may not be accurate.  The
coarse-graining  procedure here presented makes the  investigation of
the low
energy domain possible for a general static manifold. The
renormalization group analysis formulated in the Kadanoff-Wilson
scheme
controls the evolution of the new effective vertices as we scale
towards the
infrared.  In particular, by linearizing the renormalization group
transformation close to the UV fixed point, the renormalized flow can
be
explicitly studied in the improved scheme.
In the massive model the flow slows down as we move in the infrared
direction
and the infrared fixed point is recovered below the mass gap for
arbitrary
values of the conformal coupling $\xi_{R}$.

These  results have been found for a positive $\xi_{R}R$ term and it
would be
interesting to address the $\xi_{R}\leq 0$ case where an increase of
curvature
can drive the system in the spontaneously broken phase.
\acknowledgments
The author would like to thank J.Polonyi for his constant advice and
for many
enlightening conversations. The author has also benefitted from
discussions
with M.Reuter, B.-L.Hu, L.Ford, E.Calzetta, and B.Teshima.
\section{appendix}
This appendix contains some relevant details concerning the
calculations. The
one loop contribution to the local potential has been calculated by
means of
the ``generalized" $\zeta$-function \cite{hw}.  We have from
(\ref{zeta1})
\bea
&&\zeta_{n}(s)=\frac{\beta\mu^{2s}}{2\pi} \int
dp\sum_{n'=n+1}^{\infty}\sum_{l=0}^{n'}\sum_{m=-l}^{m=l}
(\omega^2(p,n'))^{-s} \nonumber\\[2 mm]
&&=\frac{\beta (\mu
a)^{2s}}{a\pi}\int_{0}^{\infty}dp\sum_{m=n+1}^{\infty}
\frac{(m+1)^2}{(p^2+(m+1)^2+\nu)^s},
\hspace{2 cm}n\geq 0\nonumber\\[2 mm]
&&=\frac{\beta(\mu
a)^{2s}}{a\sqrt{4\pi}}\frac{\Gamma(s-1/2)}
{\Gamma(s)}\sum_{m=n}^{\infty}\frac{m^2}{(m^2+\nu)^{s-1/2}},
\hspace{2 cm} n\geq 1
\label{zzrt}
\eea
where where $a^2(m^2+\frac{1}{2}\lambda\Phi^2+\xi R)-1=\nu>-1$ for
$\xi>0$. In
the last line we have set $n+1\rightarrow n$ so that the range of $n$
starts at
$n=1$ and
the sum coincides with the $\zeta$-function, extensively studied by
several
authors
\cite{hu,to,den,fo}, whose derivative in $s=0$  gives the one-loop
quantum
contribution to the effective potential in the Einstein universe. For
general
$n$, the sum in the last line can be performed with the Plana
summation formula
 \cite{ghika,ww},
\be
g(s)=\sum_{m=n}^{\infty}f(s,m) = \frac{1}{2}f(s,n)+\int_{n}^\infty
f(s,\tau)d\tau+i\int_{0}^\infty\frac{f(s,n+iz)-f(s,n-iz)}{e^{2\pi
z}-1}dx\label{zrt}
\ee
with
\be
f(s,m)=\frac{m^2}{(m^2+\nu)^{s}}.
\ee
The resulting  $g(s)$ has simple poles at $s-3/2=0,-1,...$. Taking the
derivative of  (\ref{zzrt}) in $s=0$ with $g(s)$ in (\ref{zrt}) one
obtains for
$s=0$  expression
(\ref{reno}), where
\be
\Upsilon_{n}(\Phi)=-\frac{\nu^2}{8}\ln[n+\sqrt{n^2+\nu}]
+\frac{1}{8}n\sqrt{n^2+\nu}(\nu+2n^2-4n)-{\cal
I}_{n}(\Phi)\label{fff}
\ee
and
\be
{\cal I}_{n}(\Phi)=\int_{0}^\infty\frac{4nze^{-\pi z}}{{\rm sinh }\pi
z}\frac{2z^2(\nu+5n^2)-n^2(2\nu+3n^2)-3z^4}{{\rm
Re}[(n+iz)^2\sqrt{(n+iz)^2+\nu}]}
\ee
A careful evaluation of the $\nu\gg 1$ limit for $n=1$ in expression
(\ref{fff}) shows that
\be
\Upsilon_{1}(\Phi)\sim\frac{\nu^2}{16}\ln\nu+O(1/\nu).
\ee


\end{document}